\documentclass{jpsj2}
\newtheorem{theorem}{Theorem}
\newtheorem{lemma}{Lemma}
\newcommand{\qed}{\hbox{\rule[-2pt]{3pt}{6pt}}}

\title{
Inequalities for the Local Energy of Random Ising Models
}

\author{Hidetsugu \textsc{Kitatani}$^{1}$, Hidetoshi \textsc{Nishimori}$^{2}$
and Akira \textsc{Aoki}$^{1}$}

\inst{$^{1}$Department of Electrical Engineering, Nagaoka University of
Technology, Nagaoka, Niigata 940-2188 \\
$^{2}$Department of Physics, Tokyo Institute of Technology, Oh-okayama,
Meguro-ku, Tokyo 152-8551}

\abst{
We derive a rigorous lower bound on the average local
energy for the Ising model with quenched randomness.
The result is that the lower bound is given by the average
local energy calculated in the absence of all interactions
other than the one under consideration.
The only condition for this statement to hold is that the distribution
function of the random interaction under consideration is symmetric.
All other interactions can be arbitrarily distributed including non-random cases.
A non-trivial fact is that any introduction of other interactions to the isolated
case always leads to an increase of the average local energy,
which is opposite to ferromagnetic systems where the Griffiths inequality holds.
Another inequality is proved for asymmetrically distributed interactions.
The probability for the thermal average of the local energy to be lower than that
for the isolated case takes a maximum value on the Nishimori
line as a function of the temperature.
In this sense the system is most stable on the Nishimori line.
}

\kword{spin glass, random interaction, local energy, lower bound, Nishimori line}
\begin{document}
\maketitle

\section{Introduction}

Studies of random spin systems, especially spin glasses, have a long history,
and a good amount of knowledge has been accumulated.\cite{HN1,MPV}
It has nevertheless been very difficult to establish reliable analytical results
for finite-dimensional systems, where the mean-field theory may not apply.
A notable exception is the gauge theory of spin glasses, by which the exact energy
can be calculated for a generic system on the Nishimori line (NL).\cite{HN1,HN2}
Several other rigorous results have also been derived within the same
theoretical framework.
Another class of rigorous results relevant to the present work is a set of
correlation inequalities proved for random spin systems under certain circumstances,
which are related to the existence of thermodynamic limit of various physical
quantities.\cite{CG,CG2,CMN,MNC,CL}
It is very useful and important to find further rigorous relations
to clarify the properties of random spin systems.

The purpose of the present paper is to analyze the behaviour of the local
energy of random spin systems with symmetric and asymmetric distribution
functions of interactions.
The first part of the paper concerns a generic case, not necessarily spin glasses,
and therefore we do not use the gauge theory.
We instead derive a relation to compare the average local
energy with the same quantity calculated in the absence of all other interactions.
The result is surprisingly simple that the former is always larger than or
equal to the latter.

Another inequality is derived for the spin glass problem.
We evaluate the probability that the thermal average of the local energy is
lower than the same quantity in the absence of all other interactions.
We find that this probability is a non-monotonic function of the temperature
and assumes its maximum value on the NL.

These two inequalities are quite non-trivial with possibilities to give us
hints on the detailed spin states in random systems.
We prove these two inequalities in the following two sections.
The last section is devoted to discussions.

\section{Lower bound on the local energy for symmetric distribution of randomness}

Let us first define several quantities.
We treat the Ising spin system described by the Hamiltonian
\begin{equation}
 {\cal H} = -\sum_{A\subset V}J_{A}\prod_{i\in A}S_{i}.
  \label{Hamiltonian}
\end{equation}
Here $V$ is the set of sites, and the sum over $A$ runs over all subsets
of $V$ among which interactions exist.
The number of sites in $A$ is arbitrary and may be different from
subset to subset, including the case of an external field (single-site interaction) $|A|=1$.
The lattice structure is assumed to be reflected in the choice of $A$
for which $J_A\ne 0$.

The probability distribution of the random interaction $J_A$ is denoted by $P(J_{A})$.
This distribution function is in general allowed to change from subset to subset,
including the case without randomness having $P(J_A)=\delta (J_A-J)$.

We denote the thermal average of the local energy,
$-J_A\prod_{i\in A}S_{i}$, by angular brackets as
\begin{equation}
  e_{A}(\beta,J_{A})=-\left\langle J_{A}\prod_{i \in A}S_{i}\right\rangle_{\beta},
\end{equation}
where $\beta$ is the inverse temperature $\beta =1/k_B T$ with $k_B$ being the
Boltzmann constant.
The configurational average over the distribution of randomness in interactions
will be written as $[ \cdots ]$.

As a reference energy we define the configurational and thermal average
of the local energy in the absence of all interactions other than the one
under consideration as
\begin{equation}
 e_{A}^0(\beta)=-\int_{-\infty}^{\infty}dJ_{A}P(J_{A}) J_{A} \tanh (\beta J_{A}).
\end{equation}
The `average' local energy will stand for the local energy averaged over both
the configuration of randomness and thermal fluctuations.

Our result is the following theorem.
\begin{theorem}\label{theorem1}
The average local energy is larger than or equal to $e_{A}^0(\beta)$
if the distribution of randomness is symmetric
for the interaction under consideration $P(J_A)=P(-J_A)$:
\begin{equation}
  [e_{A}(\beta,J_{A})] \geq e_{A}^0(\beta).
\end{equation}
\end{theorem}
\vspace{2mm}
{\em Remark.}\quad
There are no conditions whatsoever on the distribution functions of
interactions other than $J_A$ or the lattice structure.
Thus it is possible, for example, that all other interactions
are non-random and ferromagnetic.

To prove this theorem it is useful to bring the following result into attention.
\begin{lemma}
Let us divide the thermal average of the local energy into the trivial part
and the rest:
 \begin{equation}
  e_{A}(\beta,J_{A})=-J_{A}\tanh (\beta J_{A})+\Delta (\beta,J_{A}).
 \label{eq:lemma1}
 \end{equation}
 The first term on the right hand side is trivial because it is the
 thermal average of the local energy
 in the absence of all interactions other than $J_A$.
 Then $\Delta (\beta,J_{A})$ satisfies the following identity:
 \begin{equation}
  \Delta (\beta,-J_{A})=-\frac {\Delta (\beta,J_{A})}{1+\displaystyle
  \frac{\sinh(2\beta J_{A})}{J_{A}}\Delta (\beta,J_{A})}.
  \label{Delta_identity}
 \end{equation}
\end{lemma}
{\em Proof.}\quad
Let us denote the partition function of the system for a given configuration of
randomness as
\begin{equation}
 Z(\beta ,J_{A})=\sum_{\{S\}}\exp\left(\beta\sum_{B\subset V,B \neq
 A}J_{B}\prod_{i\in B}S_{i}+\beta J_{A}\prod_{i \in A}S_{i}\right),
\end{equation}
where we write the interaction $J_A$ separately for later convenience.
Now we evaluate the ratio of $Z(\beta ,-J_{A})$ to $Z(\beta ,J_{A})$:
\begin{eqnarray}
 \frac {Z(\beta,-J_{A})}{Z(\beta,J_{A})}
  &=&\cosh (2\beta J_{A})+\frac{\sinh(2\beta J_{A})}{J_{A}}
  e_{A}(\beta,J_{A}) \nonumber \\
   &=&1+\frac {\sinh (2\beta J_{A})}{J_{A}}\left(e_{A}(\beta,J_{A})
 +J_{A}\tanh (\beta J_{A})\right).
  \label{Z_ratio1}
\end{eqnarray}
Similarly we obtain
\begin{equation}
 \frac {Z(\beta,J_{A})}{Z(\beta,-J_{A})}
 =1+\frac {\sinh(2\beta J_{A})}{J_{A}}(e_{A}(\beta,-J_{A})
  + J_{A}\tanh (\beta J_{A})).
 \label{Z_ratio2}
\end{equation}
Since eqs. (\ref{Z_ratio1}) and (\ref{Z_ratio2}) are reciprocal to each other, we have
\begin{equation}
  1+\frac{\sinh(2\beta J_{A})}{J_{A}}\Delta (\beta,J_{A})=
  \frac {1}{\displaystyle 1+\frac{\sinh(2\beta J_{A})}{J_{A}}\Delta (\beta,-J_{A})},
\end{equation}
which leads to the desired relation (\ref{Delta_identity}).
\qed
\\
{\em Remark.}\quad
Equation (\ref{Delta_identity}) implies that $\Delta (\beta,J_A)$ and $\Delta (\beta,-J_A)$ 
have opposite signs.  Note that the denominator of eq. (\ref{Delta_identity}) is
always positive as can be verified from eq. (\ref{Z_ratio1}).
\\
{\em Proof of Theorem \ref{theorem1}.}\quad
We first rewrite the definition of the configurational average of $e_A(\beta,J_A)$,
\begin{eqnarray}
 \left[e_{A}(\beta,J_{A})\right]
  &=&\int_{-\infty}^{\infty}dJ_{A}P(J_{A})
  \left[(-J_{A}\tanh (\beta J_{A})+\Delta (\beta,J_{A})\right]^{'}
               \nonumber \\
   &=&-\int_{-\infty}^{\infty}dJ_{A}P(J_{A})
  J_{A}\tanh (\beta J_{A})+[\Delta (\beta,J_{A})]
                  \nonumber \\ 
  &=& e_{A}^0(\beta)+[\Delta (\beta,J_{A})],
\end{eqnarray}
where $[\cdots ]^{'}$ stands for the configurational average over randomness
of other interactions than $J_A$. 
For the configurational average of $\Delta (\beta, J_A)$, we have
\begin{eqnarray}
 \left[\Delta (\beta,J_{A})\right]
   &=&\int_{-\infty}^{\infty}dJ_{A}P(J_{A})[\Delta(\beta,J_{A})]^{'}
           \nonumber \\
   &=&\int_{0}^{\infty}dJ_{A}\left(P(J_{A})[\Delta(\beta,J_{A})]^{'}
    +P(-J_{A})[\Delta(\beta,-J_{A})]^{'}\right)
            \nonumber \\
   &=&\int_{0}^{\infty}dJ_{A}P(J_{A}) [\Delta(\beta,J_{A})+\Delta(\beta,-J_{A})]^{'},
   \label{Delta_int1}
\end{eqnarray}
where we have used the symmetry of the probability distribution $P(J_A)=P(-J_A)$.
According to Lemma 1, the sum of $\Delta (\beta,J_A)$ and  $\Delta (\beta,-J_A)$
in the integrand of eq. (\ref{Delta_int1}) satisfies
\begin{equation}
 \Delta(\beta,J_{A})+\Delta(\beta,-J_{A})=
   \frac {\displaystyle \frac{\sinh(2\beta J_{A})}{J_{A}}\Delta (\beta,J_{A})^{2}}{1+
   \displaystyle \frac{\sinh(2\beta J_{A})}{J_{A}}\Delta (\beta,J_{A})}.
\end{equation}
Therefore the following relation holds,
\begin{equation}
 [e_{A}(\beta,J_{A})]=e_{A}^0(\beta)
   +\int_{0}^{\infty}dJ_{A}P(J_{A})
 \left[\frac {\displaystyle \frac{\sinh(2\beta J_{A})}{J_{A}}\Delta (\beta,J_{A})^{2}}
 {1+ \displaystyle
  \frac{\sinh(2\beta J_{A})}{J_{A}}\Delta (\beta,J_{A})}\right]^{'},
  \label{e_average}
\end{equation}
from which Theorem \ref{theorem1} follows. \qed
\\
{\em Remark.}\quad
The inequality proved in Theorem 1 is the best possible one as a generic inequality
which is valid for any system.
The reason is that a generic inequality should apply to the case where only the
given $J_A$ exists and all other interactions are absent.
We therefore have to take into account the characteristics of specific systems
if we are to improve the lower bound.

For example, let us consider a triangle made of three sites,
between any two spins on which there exists the usual two-body interaction.
Each of these three spins may have interactions with spins on
other sites outside the triangle.
The two-body interactions within the triangle
are assumed to obey the same symmetric $\pm J$ distribution.
Then the configurational average of the sum of the thermal averages of
these three interaction energies
$e_{\rm TR}(\beta)$ satisfies the following inequality as shown in Appendix:
\begin{equation}
  [e_{\rm TR}(\beta)] \geq e_{\rm TR}^0(\beta),
  \label{e_inequality}
\end{equation}
where 
$e_{\rm TR}^0(\beta)$ is the configurational average of the same energy
in the absence of all other interactions,
\begin{equation}
 e_{\rm TR}^0(\beta)=-\frac {3}{2}J
   \left(\frac {\exp(2\beta J)-\exp(-2\beta J)}{\exp(2\beta J)
   +3\exp(-2\beta J)}+\frac {\exp (2\beta J)-\exp(-2\beta J)}
   {3\exp(2\beta J)+\exp (-2\beta J)}\right).
\end{equation}
This quantity satisfies the following inequality
\begin{equation}
 e_{\rm TR}^0(\beta)\ge -3J \tanh (\beta J)
\end{equation}
in consistency with Theorem \ref{theorem1}.
Equation (\ref{e_inequality}) is an improvement over Theorem \ref{theorem1},
which claims that $[e_{\rm TR}(\beta)] \geq -3J \tanh (\beta J)$,
for the special case of the triangular arrangement of three sites.

\section{Local energy for asymmetric distribution of randomness}

We next analyze the probability distribution of the local energy for asymmetric
distribution function of random interactions.

Assume that all the interactions in the system share the same distribution
function satisfying
\begin{equation}
  P(-J_{A})=\exp (-2\beta_{\rm N}J_{A})P(J_{A}),
\end{equation}
where $\beta_{\rm N}$ is a parameter corresponding to the inverse temperature
on the NL as will be exemplified below.
This constraint is known to allow us to apply the gauge theory.\cite{HN2,HN1}
For example, in the case of the $\pm J$ model
\begin{equation}
 P(J_{A})=p\delta (J_{A}-J)+(1-p)\delta (J_{A}+J)\quad\left(J>0,~p>\frac{1}{2}\right),
\end{equation}
we find
\begin{equation}
  \exp (2\beta_{\rm N} J)=\frac {p}{1-p}.
\end{equation}
For the Gaussian model
\begin{equation}
 P(J_{A})=\frac {1}{\sqrt{2\pi J^{2}}}\exp \left\{
   -\frac {(J_{A}-J_{0})^{2}}{2J^{2}}\right\}\quad (J_{0}>0),
\end{equation}
$\beta_{\rm N}$ is
\begin{equation}
  \beta_{\rm N}=\frac {J_{0}}{J^{2}}.
\end{equation}
Our result is the following theorem.
\begin{theorem}
\label{theorem2}
The system defined above satisfies the following inequality,
\begin{equation}
 \left| \left[ \frac {\Delta (\beta,J_{A})}{\mid \Delta (\beta,J_{A}) \mid}\right] \right|
  \leq -\left[ \frac {\Delta (\beta_{\rm N},J_{A})}{\mid \Delta (\beta_{\rm N},J_{A})
  \mid}\right] \label{Delta_inequality}
\end{equation}
with $\Delta (\beta,J_A)$ defined in eq. (\ref{eq:lemma1}).
\end{theorem}
\vspace{2mm}
{\em Remark.}\quad
The ratio
\begin{equation}
  -\frac {\Delta (\beta,J_{A})}{\mid \Delta (\beta,J_{A}) \mid}
\end{equation}
is $+1$ if the thermal average of the local energy $e_A(\beta,J_A)$ is lower than
the trivial value $-J_A\tanh (\beta J_A)$ and is $-1$ otherwise.
Thus the configurational average of this ratio is the difference of
$p_1$ and $p_2$, where $p_1$ is the probability
that $e_A(\beta,J_A)$ is lower than $-J_A\tanh (\beta J_A)$
and $p_2$ is for the other case.
Since $p_1+p_2=1$, the difference satisfies $p_1-p_2=2p_1-1$.
Hence the configurational average of the above ratio is a measure
of the probability $p_1$ that $e_A(\beta,J_A)$ is lower than $-J_A\tanh (\beta J_A)$.
\\
{\em Proof.}\quad
We rewrite the ratio as follows:
\begin{eqnarray}
  \left[\frac {\Delta (\beta,J_{A})}{\mid \Delta (\beta,J_{A})
  \mid}\right]
   &=&
  \int_{-\infty}^{\infty}P(J_{A})
   \left[\frac {\Delta (\beta,J_{A})}{\mid \Delta (\beta,J_{A}) \mid}\right]^{'}
                    \nonumber \\
  &=&
   \int_{-\infty}^{\infty}P(-J_{A})
   \left[\frac {\Delta (\beta,-J_{A})}{\mid \Delta (\beta,-J_{A}) \mid}\right]^{'}
                    \nonumber \\
   &=&
   \int_{-\infty}^{\infty}P(J_{A})\exp (-2\beta_{\rm N}J_{A})
   \left[\frac {\Delta (\beta,-J_{A})}{\mid \Delta (\beta,-J_{A}) \mid}\right]^{'}
                    \nonumber \\
  &=&
    \left[\exp (-2\beta_{N}J_{A}) \frac {\Delta (\beta,-J_{A})}{\mid \Delta (\beta,-J_{A})
    \mid}\right].
\end{eqnarray}
Equation (\ref{Delta_identity}) implies
\begin{equation}
 \frac {\Delta (\beta,-J_{A})}{\mid \Delta (\beta,-J_{A})\mid} =
  \frac {-\Delta (\beta,J_{A})}{\mid \Delta (\beta,J_{A})\mid}
\end{equation}
Thus, we obtain
\begin{eqnarray}
  \left[\frac {\Delta (\beta,J_{A})}{\mid \Delta (\beta,J_{A}) \mid}\right]
   &=&
  \left[\exp (-2\beta_{\rm N}J_{A})
  \frac {-\Delta (\beta,J_{A})}{\mid \Delta (\beta,J_{A}) \mid}\right]
           \nonumber \\
   &=&
   \left[\left(\cosh(2\beta_{\rm N}J_{A})-\sinh(2\beta_{\rm N}J_{A})\right)
   \frac {-\Delta (\beta,J_{A})}{\mid \Delta (\beta,J_{A}) \mid}\right].
\end{eqnarray}
According to the gauge theory,\cite{HN1,HN2} we may perform the local gauge transformation,
\begin{equation}
 S_{i}\rightarrow S_{i}\sigma_{i} \quad J_{A} \rightarrow J_{A}\prod_{i\in A} \sigma_{i},
\end{equation}
where $\sigma_{i}$ is a local gauge variable which takes
the values $\pm 1$ (arbitrarily fixed at each site).
It is useful to note that $\Delta (\beta,J_{A})$ is invariant under the above
gauge transformation.
Then, we obtain
\begin{eqnarray}
 &&\left[\frac {\Delta (\beta,J_{A})}{\mid \Delta (\beta,J_{A})
  \mid}\right]
 =
  \left[\left(\cosh(2\beta_{\rm N}J_{A})-\sinh(2\beta_{\rm N}J_{A})
 \prod_{i \in A} \sigma_{i}\right)
 \frac {-\Delta (\beta,J_{A})}{\mid \Delta (\beta,J_{A})\mid}\right]
      \nonumber \\
  &&=
  \left[\left(\cosh(2\beta_{\rm N}J_{A})-\sinh(2\beta_{\rm N}J_{A})
  \left\langle\prod_{i \in A}\sigma_{i}\right\rangle_{\beta_{\rm N}}\right)
  \frac {-\Delta (\beta,J_{A})}{\mid \Delta (\beta,J_{A}) \mid}\right]
      \nonumber \\
  &&=
  \left[\left(1+\frac{\sinh(2\beta_{\rm N}J_{A})}{J_{A}}\left(-J_{A}
  \left\langle\prod_{i \in A}\sigma_{i}\right\rangle_{\beta_{\rm N}}
  +J_{A}\tanh(\beta_{\rm N}J_{A})\right) \right)
 \frac {-\Delta (\beta,J_{A})}{\mid \Delta (\beta,J_{A}) \mid}\right].
\end{eqnarray}
Hence we have the following identity,
\begin{equation}
  \left[ \frac {\Delta (\beta,J_{A})}{\mid \Delta (\beta,J_{A})
  \mid}\right]
  =-\left[\left(1+\frac{\sinh (2\beta_{\rm N}J_{A})}{J_{A}}
  \Delta(\beta_{\rm N},J_{A})\right) \frac {\Delta (\beta,J_{A})}{\mid \Delta (\beta,J_{A})
\mid}\right],
\end{equation}
from which we find
\begin{equation}
  \left[ \frac {\Delta (\beta,J_{A})}{\mid \Delta (\beta,J_{A}) \mid}\right]
  =
  -\left[ \frac {\Delta (\beta,J_{A})}{\mid \Delta (\beta,J_{A})\mid}\right]
  -\left[\frac{\sinh (2\beta_{\rm N}J_{A})}{J_{A}}\Delta(\beta_{\rm N},J_{A})
  \frac {\Delta (\beta,J_{A})}{\mid \Delta (\beta,J_{A}) \mid}\right].
\end{equation}
We therefore have
\begin{equation}
 -\left[ \frac {\Delta (\beta,J_{A})}{\mid \Delta (\beta,J_{A})\mid}\right]=\left[
  \frac {\sinh (2\beta_{\rm N}J_{A})}{2J_{A}} \Delta(\beta_{\rm N},J_{A})
 \frac {\Delta(\beta,J_{A})}{\mid \Delta (\beta,J_{A}) \mid}\right].
\label{eq:Delta_intermediate1}
\end{equation}
When $\beta=\beta_{\rm N}$ (corresponding to the NL), we get
\begin{equation}
 -\left[ \frac {\Delta (\beta_{\rm N},J_{A})}{\mid \Delta (\beta_{\rm N},J_{A})
 \mid}\right]= \left[\frac{\sinh (2\beta_{\rm N}J_{A})}{2J_{A}}
  \mid \Delta(\beta_{\rm N},J_{A}) \mid
  \right].\label{Delta_ratio1}
\end{equation}
On the other hand, for an arbitrary temperature, we obtain the following
inequality from eq. (\ref{eq:Delta_intermediate1}),
\begin{eqnarray}
  \left| \left[ \frac {\Delta (\beta,J_{A})}{\mid \Delta (\beta,J_{A})
  \mid}\right] \right|
   &=&\left|
  \left[
   \frac {\sinh (2\beta_{\rm N}J_{A})}{2J_{A}} \Delta(\beta_{\rm N},J_{A})
   \frac {\Delta(\beta,J_{A})}{\mid \Delta (\beta,J_{A})\mid}\right] \right|
     \nonumber \\
  &\leq&\left[\frac {\sinh (2\beta_{\rm N}J_{A})}{2J_{A}}
   \mid \Delta(\beta_{\rm N},J_{A}) \mid \right]. \label{Delta_ratio2}
\end{eqnarray}
From eqs. (\ref{Delta_ratio1}) and (\ref{Delta_ratio2}),
we derive eq. (\ref{Delta_inequality}). \qed
\\
{\em Remark.}\quad
The quantity $-J_A\tanh (\beta J_A)$ is the thermal average of
the local energy in the absence of all other interactions (isolated case).
Theorem \ref{theorem2} states that
the deviation $\Delta (\beta,J_A)$ from the isolated case is more likely
to be negative on the NL than at any other temperatures.
In this sense the system is most stable on the NL than at any other
temperatures.
This observation reminds us of a similar non-monotonic behaviour
of the spin orientation as discussed previously.\cite{HN3}

\section{Summary and discussion}
We have proved two inequalities on the behaviour of the local energy for
the Ising model with random interactions.
The first inequality concerns the case of symmetric distribution function
of randomness and states that the configurational and thermal average of
the local energy is not smaller than that of the same local energy
in the absence of all other interactions.
Therefore an introduction of any interactions to the isolated case always
increases the average local energy.
This is to be contrasted to the result for ferromagnetic systems,
the Griffiths inequality,\cite{G} in which an introduction of other
interactions always lowers the thermal average of the local energy.
The random system under consideration behaves quite differently
from the conventional ferromagnetic systems.
Notice that our result is a special case of the conjecture presented by
Contucci and Lebowitz.\cite{CL}

The second inequality is for asymmetric distribution of random interactions.
The probability of decrease of the thermal average of the
local energy from the isolated case is a non-monotonic function of the temperature
and assumes the largest value on the NL.
This is again a somewhat counter-intuitive result.
Note that the average of $\Delta (\beta, J_A)$ itself, not the sign of it,
vanishes on the NL.\cite{HN1,HN2}
The sign of $\Delta (\beta, J_A)$ may be positive or negative depending
on the configuration of randomness.
If we focus ourselves only on the sign of $\Delta (\beta, J_A)$,
ignoring the magnitude, we find that
the probability for this quantity to be negative is largest on the NL.
The NL therefore occupies a special position in the phase diagram
from the view point of spin configurations.
Further clarifications are necessary.
\section*{Acknowledgement}
We thank Pierluigi Contucci for useful comments.
The work of HN was  supported by CREST, JST and by the Grant-in-Aid
for Scientific Research on Priority Area `Deepening and Expansion of
Statistical Mechanical Informatics' by the Ministry of Education,
Culture, Sports, Science and Technology.

\appendix
\section{Derivation of eq. (\ref{e_inequality})}

In this appendix we explain the derivation of eq. (\ref{e_inequality}).
Let us consider the energy of a three-spin system
consisting of spins $S_{1}$, $S_{2}$ and $S_{3}$, which
interact with each other by three two-body interactions.
We number the configurations of three interactions,
$\{J_{12},J_{23},J_{31}\}$, namely, $\{J,J,J\}$, $\{J,-J,-J\}$,
$\{-J,J,-J\}$, $\{-J,-J,J\}$, $\{-J,-J,-J\}$, $\{-J,J,J\}$,
$\{J,-J,J\}$ and $\{J,J,-J\}$ as configuration $1$ to configuration $8$.
Note that configurations 1 to 4 are unfrustrated while 5 to 8 are frustrated.
We denote the partition function for configuration $n$ as $Z(\beta,n)$ and
the thermal average of the energy of a bond $\{ij\}$ as $e_{ij}(\beta,n)$.
Note that the interactions with spins outside the triangle are
arbitrarily fixed.

It is convenient to evaluate the ratio of two configurations:
\begin{eqnarray}
 \frac {Z(\beta,2)}{Z(\beta,1)}
  &=&
 \left\langle (\cosh (2\beta J)-\sinh (2\beta J)S_{2}S_{3})
  (\cosh (2\beta J)-\sinh (2\beta J)S_{3}S_{1})\right\rangle_{\beta,1}
          \nonumber \\
  &=&
   \cosh^{2}(2\beta J)+\cosh(2\beta J)\sinh (2\beta J)
   \frac{e_{23}(\beta,1)+e_{31}(\beta,1)}{J}
          \nonumber \\
  & &-\sinh^{2}(2\beta J)\frac{e_{12}(\beta,1)}{J},
  \label{Z12}
\end{eqnarray}
where $\langle \cdots \rangle_{\beta,1}$ denotes the
thermal average at inverse temperature $\beta$ under the bond configuration  1.

Now we note the trivial inequality,
\begin{equation}
 \sum_{i=1}^{4}\sum_{j<i}
 \left(\frac {Z(\beta,j)}{Z(\beta,i)}+\frac {Z(\beta,i)}{Z(\beta,j)} \right) \geq 12.
 \label{Zsum}
\end{equation}
Substituting the results obtained by repeating the procedure to derive eq. (\ref{Z12})
into eq. (\ref{Zsum}), we find
\begin{equation}
 12\cosh^{2}(2\beta J)+2\cosh(2\beta J)\sinh (2\beta J)
  \sum_{i=1}^{4}\frac{e_{\rm TR}(\beta,i)}{J}-\sinh^{2}(2\beta J)\sum_{i=1}^{4}
  \frac{e_{\rm TR}(\beta,i)}{J} \geq 12,
  \label{Zsum_2}
\end{equation}
where
\begin{equation}
  e_{\rm TR}(\beta,i)=e_{12}(\beta,i)+e_{23}(\beta,i)+e_{31}(\beta,i).
\end{equation}
From eq. (\ref{Zsum_2}), it follows that
\begin{equation}
  \frac {1}{12}\sum_{i=1}^{4}e_{\rm TR}(\beta,i) \geq -J
  \frac{\exp (2\beta J)-\exp (-2\beta J)}{\exp (2\beta J)+3\exp (-2\beta J)},
  \label{ebound}
\end{equation}
where the right-hand side is equal to
the energy of a non-frustrated triangular cluster in isolation.
Similarly, it holds that
\begin{eqnarray}
 \frac {Z(\beta,6)}{Z(\beta,5)}
  &=&
  \left\langle (\cosh (2\beta J)+\sinh (2\beta J)S_{2}S_{3})
  (\cosh (2\beta J)+\sinh (2\beta J)S_{3}S_{1})\right\rangle_{\beta,5}
              \nonumber \\
  &=&
  \cosh^{2}(2\beta J)+\cosh(2\beta J)\sinh (2\beta J)
  \frac{e_{23}(\beta,5)+e_{31}(\beta,5)}{J}
         \nonumber \\
  & &+\sinh^{2}(2\beta J)\frac{e_{12}(\beta,5)}{J},
\end{eqnarray}
where $\langle \cdots \rangle_{\beta,5}$ denotes the thermal average at inverse
temperature $\beta$ in the bond configuration 5.
By the same procedure as above, we obtain
\begin{equation}
 12\cosh^{2}(2\beta J)+2\cosh(2\beta J)\sinh (2\beta J)
  \sum_{i=5}^{8}\frac{e_{\rm TR}(\beta,i)}{J}+\sinh^{2}(2\beta J)\sum_{i=5}^{8}
  \frac{e_{\rm TR}(\beta,i)}{J} \geq 12.
\end{equation}
Consequently,
\begin{equation}
  \frac {1}{12}\sum_{i=5}^{8}e_{\rm TR}(\beta,i) \geq -J
  \frac{\exp (2\beta J)-\exp (-2\beta J)}{3\exp (2\beta J)+\exp (-2\beta J)},
\end{equation}
where the right-hand side is equal to the
energy of a frustrated triangular cluster in isolation.
Finally, we obtain that
\begin{eqnarray}
 [e_{\rm TR}(\beta)]
  &=&
   \frac {1}{8}\sum_{i=1}^{8}[e_{\rm TR}(\beta,i)]^{'}
          \nonumber \\
  &\geq&
 -\frac {3}{2}J\left(
   \frac{\exp (2\beta J)-\exp (-2\beta J)}{\exp (2\beta J)+3\exp (-2\beta J)}
  +\frac{\exp (2\beta J)-\exp (-2\beta J)}{3\exp (2\beta J)+\exp (-2\beta J)}
   \right).
\end{eqnarray}
This is eq. (\ref{e_inequality}).



\begin{thebibliography}{99} 
\bibitem{HN1} H. Nishimori: {\em Statistical Physics of Spin Glasses and Information
 Processing: An Introduction}, Oxford University Press (Oxford, 2001).
\bibitem{MPV} M. M\'ezard, G. Parisi and M. A. Virasoro: {\em Spin Glass Theory
 and Beyond}, World Scientific (Singapore, 1987).
\bibitem{HN2} H. Nishimori: Prog. Theor. Phys. {\bf 66} (1981) 1169.
\bibitem{CG} P. Contucci and S. Graffi: J. Stat. Phys. {\bf 115} (2004) 581.
\bibitem{CG2} P. Contucci and S. Graffi: Commun. Math. Phys. {\bf 248} (2004) 207.
\bibitem{CMN} P. Contucci, S. Morita and H. Nishimori: J. Stat. Phys. {\bf 122} (2006) 303.
\bibitem{MNC} S. Morita, H. Nishimori and P. Contucci:
 J. Phys. A: Math. Gen. {\bf 37} (2004) L203.
\bibitem{CL} P. Contucci and J. L. Lebowitz: cond-mat/0612371.
\bibitem{HN3} H. Nishimori: J. Phys. Soc. Jpn. {\bf 62} (1993) 2973.
\bibitem{G} R. B. Griffiths: J. Math. Phys. {\bf 8} (1967) 478, 484.
\end{thebibliography}
\end{document}